\documentclass[rspublic, superscriptaddress, showpacs, notitlepage, twocolumn]{revtex4-1}
\usepackage{amsmath, amsfonts}
\usepackage{caption}
\usepackage{float}
\usepackage{amssymb}
\usepackage{color}
\usepackage[usenames,dvipsnames]{xcolor}
\usepackage{graphicx}
\usepackage{bm}
\usepackage{subfigure}
\usepackage{amssymb}
\usepackage{verbatim}
\usepackage{amssymb}
\usepackage{enumerate}
\usepackage{cleveref}

\usepackage{amsmath, amsfonts}
\usepackage{caption}
\usepackage{float}
\usepackage{amssymb}
\usepackage{color}
\usepackage[usenames,dvipsnames]{xcolor}
\usepackage{graphicx}
\usepackage{bm}
\usepackage{subfigure}
\usepackage{amssymb}
\usepackage{verbatim}
\usepackage{amssymb}
\usepackage{enumerate}
\usepackage{cleveref}

\newcommand{\quotes}[1]{``#1''}
\usepackage{soul}

\def \samsapprox{\ensuremath{\sim}}
\def \CO2{\ensuremath{\text{CO}_{2}}}
\newcommand{\B}[1]{ { #1}}

\begin{document}

\title{Termite mounds harness diurnal temperature oscillations for ventilation}

\author{Hunter King}
\affiliation{H.K. and S.A.O. contributed equally to this work. }
\affiliation{School of Engineering and Applied Sciences, Harvard University, Cambridge, MA, 02138, USA}
\author{Samuel A. Ocko}
\affiliation{H.K. and S.A.O. contributed equally to this work. }
\affiliation{Department of Physics, Massachusetts Institute of Technology, Cambridge, MA, 02139, USA}
\author{L. Mahadevan}\affiliation{Department of Physics, Massachusetts Institute of Technology, Cambridge, MA, 02139, USA}

\begin{abstract}
Many species of millimetric fungus-harvesting termites collectively build uninhabited, massive mound structures enclosing a network of broad tunnels which protrude from the ground meters above their subterranean nests. It is widely accepted that the purpose of these mounds is to give the colony a controlled micro-climate in which to raise fungus and brood by managing  heat, humidity, and respiratory gas exchange. 
While different hypotheses such as steady and fluctuating external wind and internal metabolic heating have been proposed for ventilating the mound,  the absence of direct in-situ measurement of internal air flows has precluded a definitive mechanism for this critical physiological function. By measuring diurnal variations in flow through the surface conduits of the mounds of the species {\it Odontotermes obesus}, we show that a simple combination of geometry, heterogeneous thermal mass and porosity allows the mounds to use diurnal ambient temperature oscillations for ventilation. In particular, the thin outer flute-like conduits heat up rapidly during the day relative to the deeper chimneys, pushing air up the flutes and down the chimney in a closed convection cell, with the converse situation at night. These cyclic flows in the mound flush out $\text{CO}_2$ from the nest and ventilate the colony, in a novel example of deriving useful work from thermal oscillations.
\end{abstract}

\maketitle

Many social insects which live in dense colonies \cite{leaf-CO2,bees-CO2} face the problem of
keeping temperature, respiratory gas, and moisture levels within tolerable ranges.
They solve this problem by using naturally available {structures} or building {their own} nests, mounds {or} bivouacs \cite{animal-architect}. A particularly impressive example of insect architecture is found in \B{fungus-cultivating termites of the subfamily Macrotermitinae,}
{individually only a few mm in body length},
that are well-known for their ability to build massive, complex structures \cite{korb-book,scott-architecture} without central decision-making authority \cite{scott-swarm}. The resulting structure includes a subterranean nest containing brood and symbiotic fungus, and a mound {extending $\sim$1-2m above ground}, which is primarily entered for construction and repair, but otherwise relatively uninhabited. 
The mound contains conduits that are many times larger than a termite \cite{scott-architecture}, and viewed widely as a means to ventilate the nest \cite{scott-extended}.
However, the mechanism by which it works {continues to be debated} \cite{scott, korb, korb-ventilation, luscher}.

Ventilation necessarily involves two steps:  transport of gas from underground metabolic sources to the mound surface, and transfer of gas across the porous exterior walls with the environment. While diffusion can equilibrate gradients across the mound surface \cite{ref-diff}, it does not suffice to transport gas between nest and surface \cite{Note1}. Thus, ventilation must rely on bulk flow inside the mound. Previous studies of  mound-building termites have suggested either thermal buoyancy or external wind as possible drivers, making a further distinction between steady (eg. metabolic driving \cite{luscher}, steady wind) and transient (eg. diurnal driving\cite{korb,korb-ventilation}, turbulent wind\cite{scott}) sources.  However, {the technical difficulties} of direct in-situ measurements of air flow in an intact mound and its correlation with internal and external environmental conditions has precluded differentiating between any of these hypotheses. Here, we use both structural and dynamic measurements to resolve this question by focusing on the mounds of \B{{\it Odontotermes obesus} (Termitidae, Macrotermitinae), which is common} 
in southern Asia in a variety of habitats\cite{termite-ref}.

In Fig. 1(a), we show the external geometry of a typical {\it Odontotermes obesus} mound, with {its characteristic} buttress-like structures (\quotes{flutes}) which extend radially from the center (Fig. 1(b)).   
\begin{figure*}
\centering
\includegraphics[height=0.8\columnwidth]{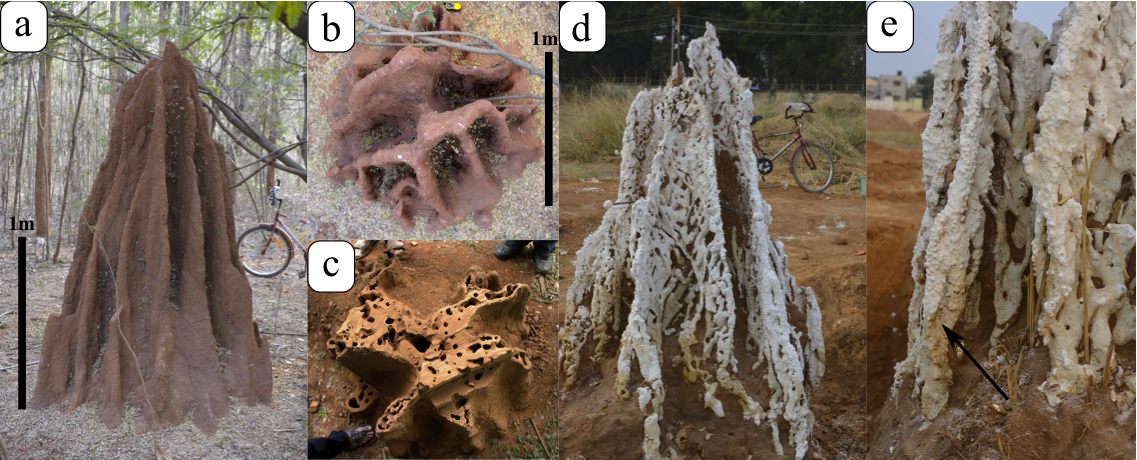}
\caption{Mounds of \emph{Odontotermes obesus} {viewed from (a) the side, (b) top and by (c) cross section. Filling the mound with gypsum, letting it set, and washing away the original material reveals the interior volume (white regions) as a continuous network of conduits, shown in (d). Endocast of characteristic vertical conduit in which flow measurements were performed, near ground level, toward the end of flutes, indicated by the arrow (e).}
}
\end{figure*}
The internal structure of the mound can be visualized using by either making a horizontal cut (Fig. 1(c)) or {endocasting} (Fig. 1(d)). Both approaches show the basic design motif of a large central chimney with many surface conduits in the flutes; all conduits are larger than termites, most are vertically oriented, and well connected \cite{Note2}.  This macro-porous structure can admit bulk internal flow and thus could serve as an external lung for the symbiotic termite-fungus colony.

To understand how the mound interacts with the environment, we first note that the walls are made of densely deposited granules of clay soil, forming a material with high porosity ({37$-$47\% air, by volume}; \cite{tejas})
, and small average pore diameter {($\sim 5 \mu m$, roughly the mean particle size)}. Indeed, healthy mounds have no visible holes to the exterior, and repairs are quickly made if the surface is breached.  The high porosity means that the mound walls provide little resistance to diffusive transport of gases along concentration gradients. However, the small pore size makes the mound very resistant to pressure driven bulk flow across its thickness. Thus, the mound surface behaves like a breathable windbreaker. Finally, the low wind speeds observed around the termite mounds of $\sim 0-5 m/s$ {implies} that they are not capable of creating significant bulk flow across the wall, effectively ruling out {wind} as the primary driving source.


Within the mound, a range of indirect measurements of $ \text{CO}_2$ concentration, local temperature, condensation, and tracer gas pulse chase\cite{luscher, scott, korb, korb-ventilation, darlington-metabolic} show the presence of transport and mixing. However,  a complete understanding of the {driving} mechanism behind these processes requires  direct measurements of flow inside the mound. This is difficult for several reasons.  First, the mound is opaque, so that any instrument must be at least partly intrusive.  Second, expected flows are small ($\sim cm /s$), outside the operating range of commercial sensors, requiring a custom-engineered device.  Third, because conduits are vertical, devices relying on heat dissipation, or larger, high heat capacity setups can generate their own buoyancy-driven (and geometry-dependent) flows, making measurements ambiguous\cite{loos}. Finally, and most importantly, the mound environment is hostile and dynamic.  Termites tend to attack and deposit sticky construction material on any foreign object, often within 10 minutes of entry.  If one inserts a sensor even briefly, termites continue construction for hours, effectively changing the geometry and hence the flow in the vicinity of the sensor. 


To measure airflow directly, we designed and built a directional flow sensor composed of three linearly arranged glass bead thermistors, exposed to the air (see appendix).  A brief pulse of current through the center bead creates a tiny bolus of warm air, which diffuses outward and is measured in either neighboring bead  \cite{Note3}. Directional flow along the axis of the beads biases this diffusion, and is quantified by the ratio of the maximum response on each bead, measured as a temperature-dependent resistance. 
In a roughly roughly conduit-sized vertical tube, this resistance-change metric depends linearly on flow velocity, with a slight upward bias due to thermal buoyancy. This allows us to measure both flow speed and direction locally.
The symmetry of the probe allows for independent calibration and measurement in two orientations by rotating by 180$^o$ (arbitrarily {labeled} '{upward}' and '{downward}')(see appendix).


In live mounds, the sensor was placed in a surface conduit at the base of a flute for $\lesssim$ 5 minutes at a time to avoid termite attacks which damage the sensors. 
For a self-check, the sensor was rotated in place,
such that a given reading could be compared on both 'upward' and 'downward' calibration curves. We also measured the flow inside an abandoned (\quotes{dead}), unweathered mound that provided an opportunity for long-term monitoring without having termites damaging the sensors.
Simultaneous complementary measurements of temperature in flutes and the center were taken.  To measure the concentrations of $ \text{CO}_2$, a metabolic product, a tube was inserted into the nest; in one mound in the center slightly below ground and another in the chimney at $\approx $1.5 m above. Gas concentration measurements were made every 15 minutes by drawing a small volume of air through an optical sensor from the two locations for most of one uninterrupted 24 hour cycle.

\begin{figure}[t]
\centering
\includegraphics[width=.95\columnwidth]{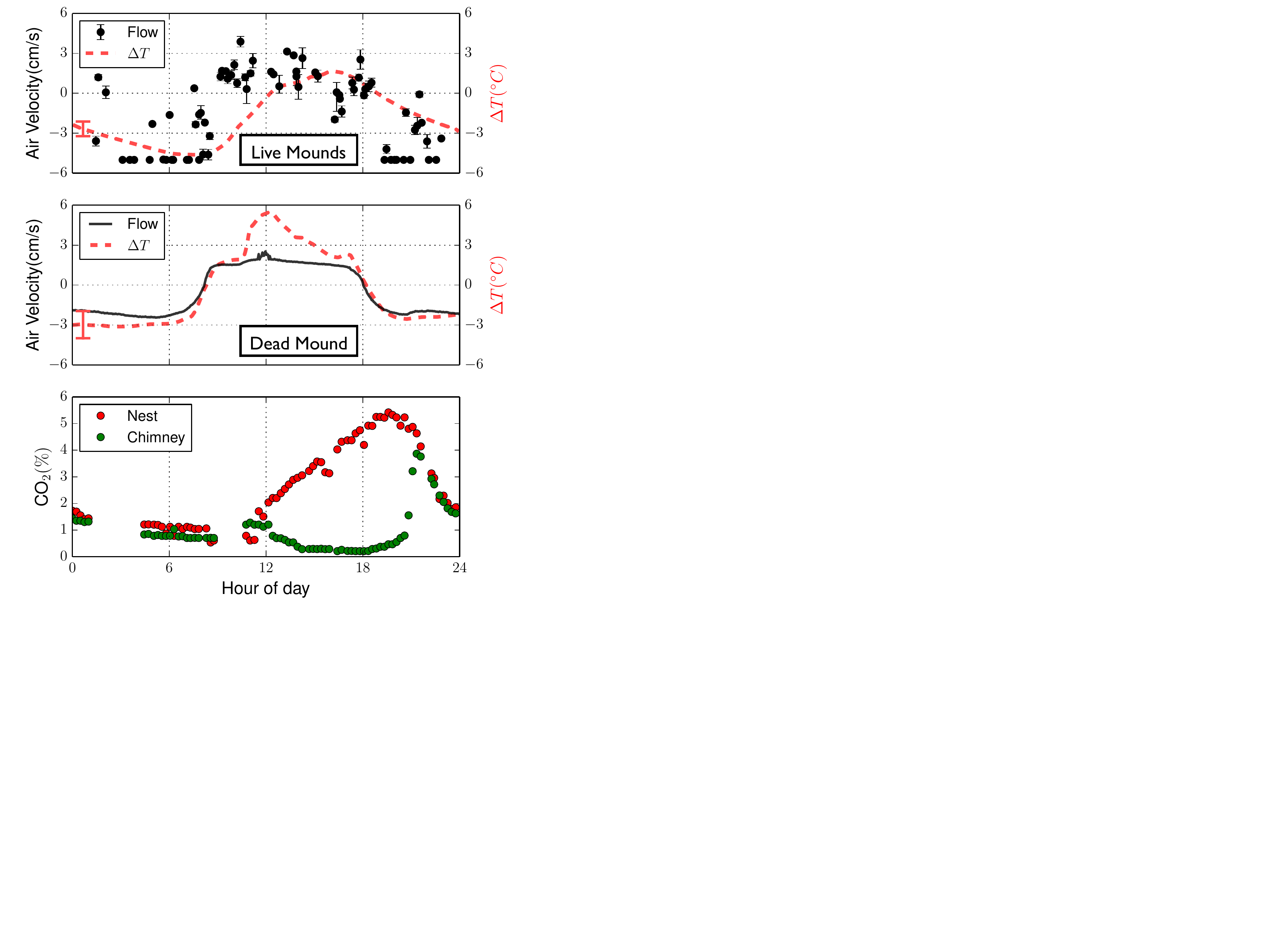}
\caption{Diurnal temperature and flow profiles show diurnal oscillations. (Top) Scatter plot of air velocity in individual flutes of 25 different live mounds ($\bullet$). Error bars represent deviation between 'upward' and 'downward' $\approx$ 1.5 minute flow measurements. The dashed red line is the average difference between temperatures measured in 4 flutes and the center (at a similar height), {$\Delta T$}, in a sample live mound(Representative error bar shown at left). (Middle) Corresponding flow and $\Delta T$, continuously measured in the abandoned mound. (Bottom) $\CO2$ schedule in the nest (${\color{red}\bullet}$) and the chimney 1.5m above (${\color{green}\bullet}$), measured over one cycle in a live mound.}
\end{figure}


Nearly all the 25 mounds that were instrumented were in a forest with little direct sunlight.  In Fig. 2(a), we show flow measurements in 78 individual flutes of these mounds as a function of time of day. We see a clear trend of slight upward (positive) flow in the flutes during the day, and significant downward (negative) flow at night.  The data saturates for many night values, as the flow speed was larger than our range of reliable calibration (see appendix).  In Fig. 2 (b), we show the flow rate for a sample flute in the abandoned mound.  Notably, it follows the same trend seen in live mounds, but the flow speeds at night are not nearly as large as for the live mounds.  For both live and dead mounds, we also show the difference in temperature measured between the flutes and center, $\Delta T=T_{flute}-T_{center}$, and see that it varies in a manner consistent with the respective flow pattern. This rules out metabolic heating \cite{luscher} as a central mechanism{, since a dead mound shows the same gradients and flows as a live one.}

In  Fig. 2(c) we show that the {accumulation of} respiratory gases also follows a diurnal cycle, with  two functioning states.   During the day, when flows are relatively small, $ \text{CO}_2$ gradually builds up to nearly 6$\%$ in the nest, and drops to a fraction of $1\%$ in the chimney.  At night, when convective flows are large, $ \text{CO}_2$ levels remain relatively low everywhere. While it is surprising that these termites allow for such large periodic accumulations of $ \text{CO}_2$,  similar tolerance has been observed in ant colonies\cite{mangrove-ant}.


In addition to measuring these slowly varying flows, we used our flow sensors in a different operating mode to also measure short-lived transient flows similar to {those of} earlier reports\cite{loos,scott}. This requires a different heating protocol, wherein the center bead is constantly heated, so that small fluctuations in flow lead to {antisymmetric} responses from the outer beads. However, the steady heating leads to a trade-off, as thermally induced buoyancy hinders our ability to interpret absolute flow rate.  Our measurements found that transients are at most only a small fraction ($\approx 1mm /s$) of average flow speeds under normal conditions, and were not induced by applying steady or pulsing wind from a powerful fan just outside the termite mound.  Pulse chase experiments on these mounds, in which combustible tracer gas is released in one location in the mound and measured in another, gave an estimate of gas transport speed (not necessarily the same as flow speed) of the same order ($\approx cm /s$), which indicates that our internally measured average flow is the dominant means of gas transport and mixing with no real role for {wind-induced} flows. Furthermore, temperature measurements in the center at different heights showed that the nest is almost always the coolest part of the mound central axis, additional evidence against the importance metabolic heating\cite{luscher} (see appendix).

\begin{figure}[t]
\centering
\includegraphics[width=0.95\columnwidth]{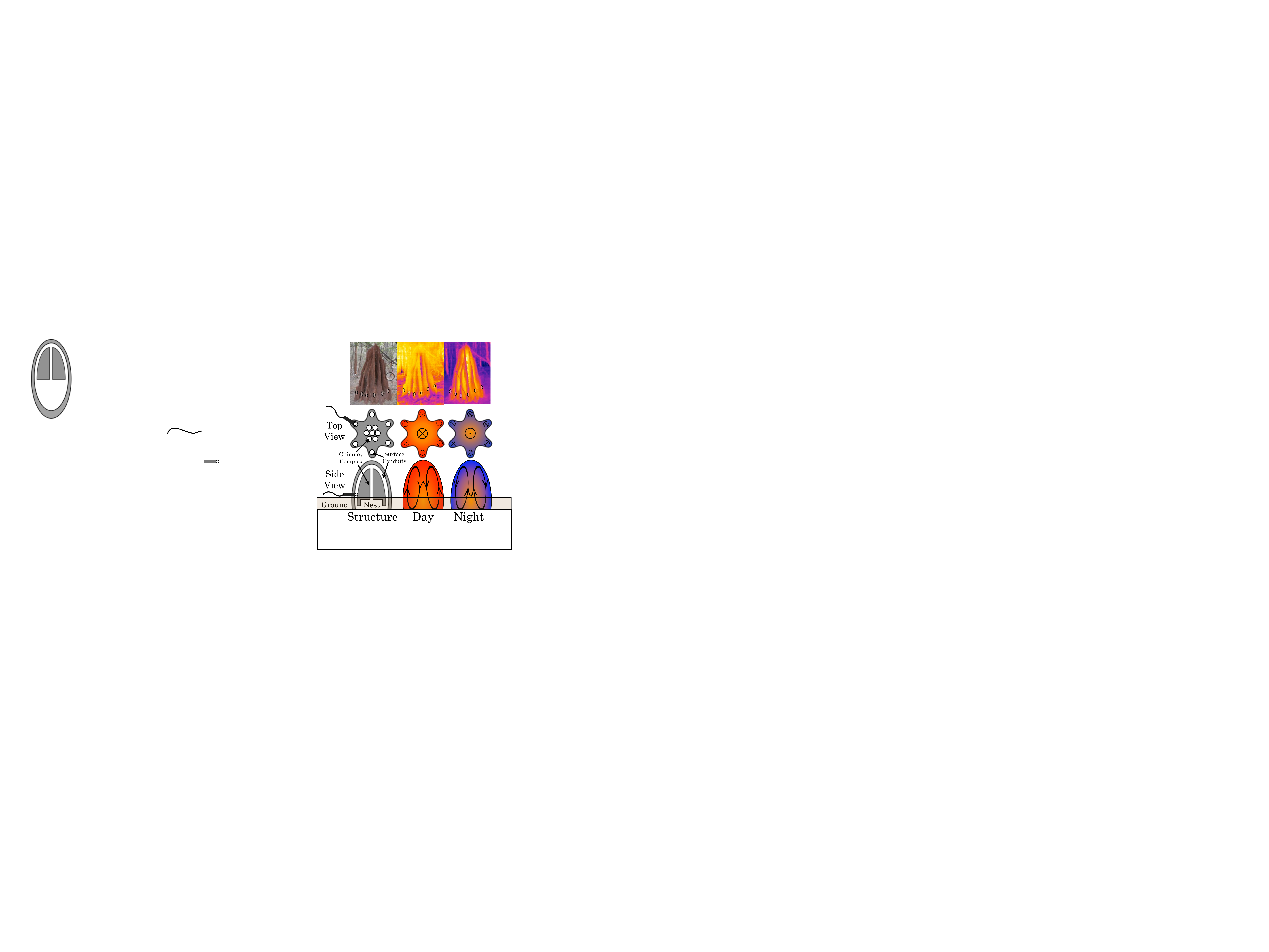}
\caption{(Top) Thermal images of the mound in 1(a), during the day and night qualitatively show an inversion of the difference between flute and nook surface temperature. Bases of flutes are marked with ovals to guide the eye. (Middle and Bottom) Mechanism of convective flow illustrated by schematic of the inverting modes of ventilation in a simplified geometry. Vertical conduits in each of the flutes are connected at top and in the subterranean nest to the vertical chimney complex. This connectivity allows for alternating convective flows driven by the inverting thermal gradient between the massive, thermally damped, center and the exposed, slender flutes, which quickly heat during the day, and cool during the night.}
\end{figure}

Taken together, these results point strongly to the idea that diurnally driven thermal gradients drive air within the mound, facilitating transport of respiratory gases.  Observations of the well connectedness of the mound and the impermeability of the external walls imply that flow in the center of the mound, to obey continuity, must move in the vertical direction opposite that of the flutes.  When the flutes are warmer than the interior, air flows up in the flutes, pushing down cooler air in the chimney.  The opposite occurs when the gradient is reversed at night \cite{Note4}(Fig. 3.).

This model predicts flow speeds comparable to those observed in the mound (see appendix), and is consistent with the quick uptake and gradual decline of $ \text{CO}_2$ measured in the chimney; with the evening temperature inversion, convection begins to push rich nest air up the chimney before diffusion across the surface gradually releases $ \text{CO}_2$ from the increasingly mixed mound air. This forcing mechanism is inherently transient; if the system ever came to equilibrium and the gradient disappeared, ventilation would stop.

{It has long been though that animal-build structures, spectacularly exemplified by \B{termite mounds}, maintain homeostatic microhabitats that allow for exchange of matter and energy with the external environment while buffering against strong external fluctuations.  Our study quantifies this by showing how a collectively-built termite mound harnesses natural temperature oscillations to facilitate collective respiration.}
The radiator-like architecture of the structure facilitates a large thermal gradient between {the} insulated chimney and exposed flutes.
The mound harnesses this gradient by creating a closed flow circuit which straddles it, promoting circulation, and flushing the nest of $ \text{CO}_2$. 
Although our data comes entirely from one termite species, the transport mechanism described here is very generic and is likely dominant in similarly massive mounds with no exterior holes that are found around the globe in a range of climates.
{A natural question that our study raises is that of the rules that lead to decentralized construction of a reliably functioning mound.
While the insect behavior that leads to construction of these mounds is not well understood, it is likely that feedback cues are important. The knowledge of the internal airflows and transport mechanisms might allow us to get a window into these feedbacks and thus serve as a step towards understanding mound morphogenesis and collective decision making.}

The {swarm-built structure} described here demonstrates how work can be derived, through architecture, from the fluctuations of an intensive environmental parameter \B{--} {a qualitatively different strategy than that of most} human engineering that typically \cite{clock} extracts work from unidirectional flow of heat or matter. Perhaps this might serve to inspire the design of similarly passive, sustainable human architecture\cite{scott-rupert,biomimetic}.

\begin{acknowledgments}
{This work was supported by the Human Frontiers Science Program and the Henry W. Kendall physics fellowship (S.O.). We thank J. S. Turner for inspiring this project as well as first suggesting the route of direct air flow measurements. We thank R. Soar, P. Bardunias and S. Sane for useful conversations, P. Prasad, P. Sharma and A. Vats for help with field work, J. Weaver for help with 3D printing and J. MacArthur and S. Mishra for advice with electronics. We also thank our anonymous referees for helpful feedback. }
\end{acknowledgments}

\clearpage
\pagebreak

\onecolumngrid
\linespread{1.} 

\appendix


\section{Experimental procedures}

\subsection{Steady flow measurement}
In an area within walking distance of the campus of NCBS Bangalore, India, 25 mounds were chosen which appeared sufficiently developed ($\gtrsim$ 1m tall) and intact.  All mounds were located in at least partial shade, but received direct sunlight intermittently during the day \cite{Note5}.
The scatter plot of steady flow (Fig. 2 (a)) represents individual measurements of the flow in the large conduits found at the end of each available flute.

In order to place the probe, a hole was manually cut with a hole saw fixed to a steel rod, usually to a depth of about 1-3cm before breaking into the conduit.  With a finger, the appropriate positioning and orientation of the sensor was determined.  Occasionally, the cavity was considered too narrow ($\lesssim$ 3 cm) diameter), or the hole entered with an inappropriate angle, in which case it was sealed with putty and another attempt was made.

Because internal reconstruction, involving many termites and wet mud, continues long after any hole is made, the same flute was never  measured twice.  In a round of measurements, flows in 1-3 (out of approximately seven available) flutes of each of \samsapprox{8} nearby mounds were measured, a process which took several hours.  Each of the 25 mounds was visited 2-3 times to utilize unmeasured flutes, but deliberately at different times of day to avoid possible correlations between mound location and flow pattern.

An individual measurement of flow was taken for $\sim$2 minutes in each orientation, such that the response curves from which the metric and then flow are calculated are averaged over many pulses ($\sim$6 pulses per minute).  The error bar of each constant flow velocity measurement in Fig. 2 (a) indicates the deviation in measured flow between orientations. Care was taken to ensure that the probe temperature remained close to the interior flute temperature, and no long-term drift in measured velocity was observed as the probe equilibrated.
Periodically between measurements, the sensor was tested in the same apparatus to check that it remained calibrated, especially when thermistors were damaged or dirtied by termites and needed to be cleaned.

The dead mound referenced in the text was identified as such because no repairs were made upon cutting holes for the sensor.  As it was also intact (there were no signs that erosion had yet exposed any of the interior cavities) and within reach of electricity, it was possible to make continuous flow measurements which could be compared with the brief measurements for live mounds.

\subsection{Geometry and sources of error in flow measurement}

In situ measurements take place in a complex geometry, and {the width, shape, and surrounding features can be highly variable. This can lead to significant variation in local velocities, even causing some local velocities to go against the average trend; this is a generic feature of flow through disordered, porous media \cite{SujitPorous,NMRAndNumericalPorousScatterPaper}}. In addition, the width, shape, and impedance of a channel are different than in our calibration setup, and the position of the probe within a channel could not be exactly known.  These factors are most likely the dominant source of error for any given measurement in the field, either over- or underestimating the flow in a particular conduit.  This error, though potentially as large as a factor of $\sim$2, is reflective of the natural variation in mound geometry, is not correlated to any other parameter, and cannot mistake the direction of flow, such that the trend in average flow remains unambiguous.

\subsection{Temperature measurements}

Temperatures in the dead mound reported in Fig. 2 (b) in the main text were obtained by implanting iButtons (DS1921G, 
Maxim) into the mound using the hole saw and closing the openings with wet mud.  2 iButtons were placed in flutes at the same location where flow data had been acquired.  Another 2 iButtons were placed 5-10 cm below the surface, in the nooks between flutes, such that they were located roughly in the periphery of the central chimney.  $\Delta T$, as shown in Fig. 2, was calculated as the temperature from the measured flute minus the average temperature measured by the centrally placed iButtons.  The raw data was slightly smoothed before taking the difference, to reduce distracting jumps in data from the iButtons, which have a thermal resolution of 0.5 C.

In the a large healthy mound, digital temperature/humidity sensors (SHT11, Sensiron) were implanted at different heights near the central axis.  Screened windows protected the sensors from direct contact by termites and building material, while remaining coupled to the interior environment.  The sensors and Arduino were powered with a high capacity 12V lead acid battery and they recorded temperature for approximately two days.  iButtons were placed in the bases of flutes in four sides of the same mound.  Temperature differences reported in Fig. 2 (a) were calculated from the average of flute temperatures and central axis temperature at the corresponding height.

\begin{figure}[t]
\centering
\includegraphics[width=\columnwidth]{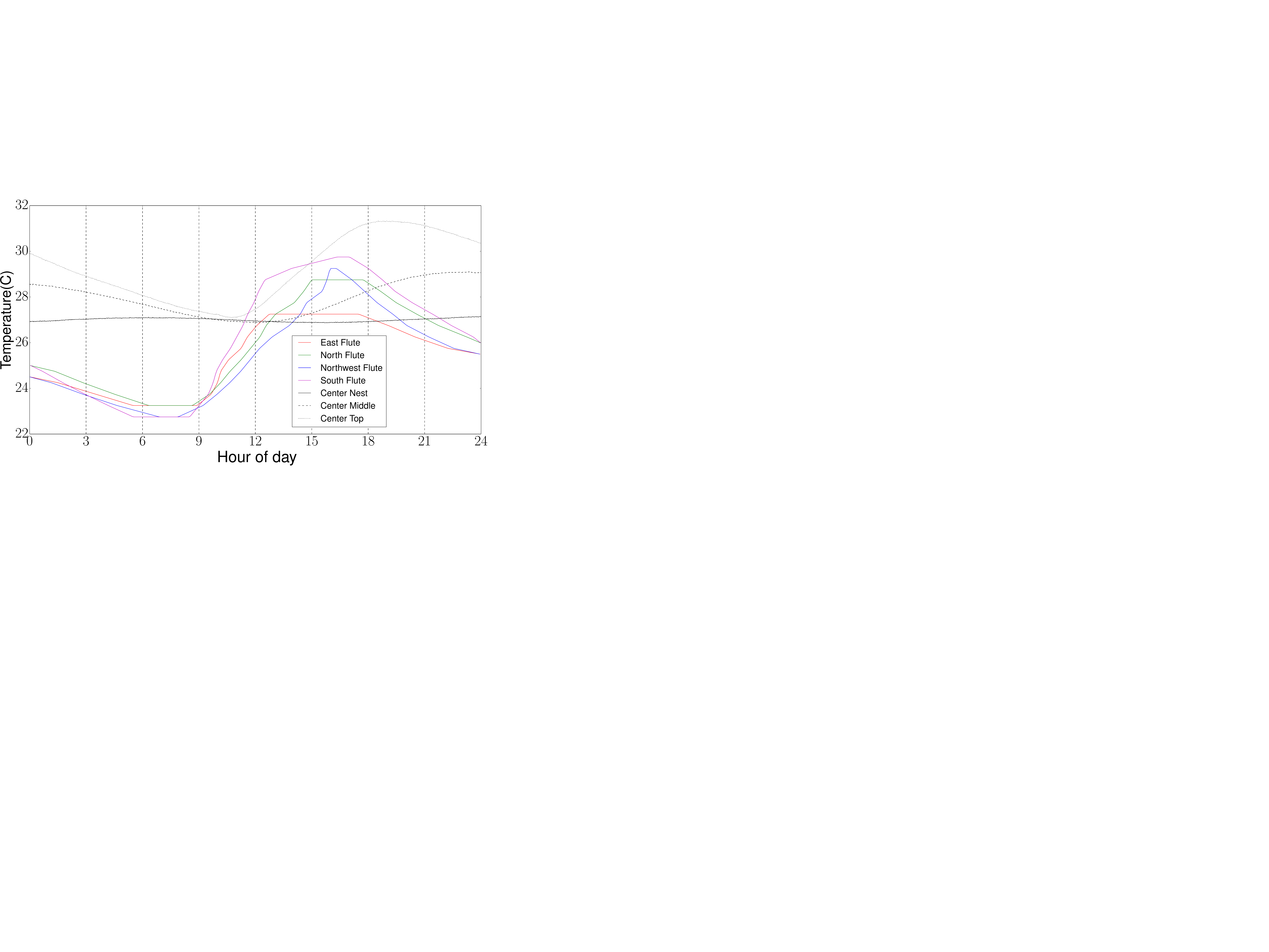}
\caption{Temperatures along the center of healthy, $\sim$2m tall, mound at three heights: \quotes{nest} ($\sim$30cm below ground), \quotes{middle} ($\sim$50cm above ground) and \quotes{top} ($\sim$130cm above ground), and in bases of flutes at 4 cardinal directions.  {Error in values along the center is $(\pm 0.4 ^{\circ}C)$, and $(\pm 1 ^{\circ}C)$ in the flutes.} The independence of behavior on cardinal direction shows direct solar heating is not of primary importance.}
\label{SD3}
\end{figure}

The temperature differences between interior and flutes shown in Fig.~\ref{SD3} are significantly larger at night than during the day. 
This and/or the vertical asymmetry of the convective cell (in that the flutes are closer to the top of the convective cell) might be responsible for the observed asymmetry in flow magnitudes between night and day.

\subsection{{Permeability and Diffusibility}}
\begin{figure}
\centering
\includegraphics[height=.45\columnwidth]{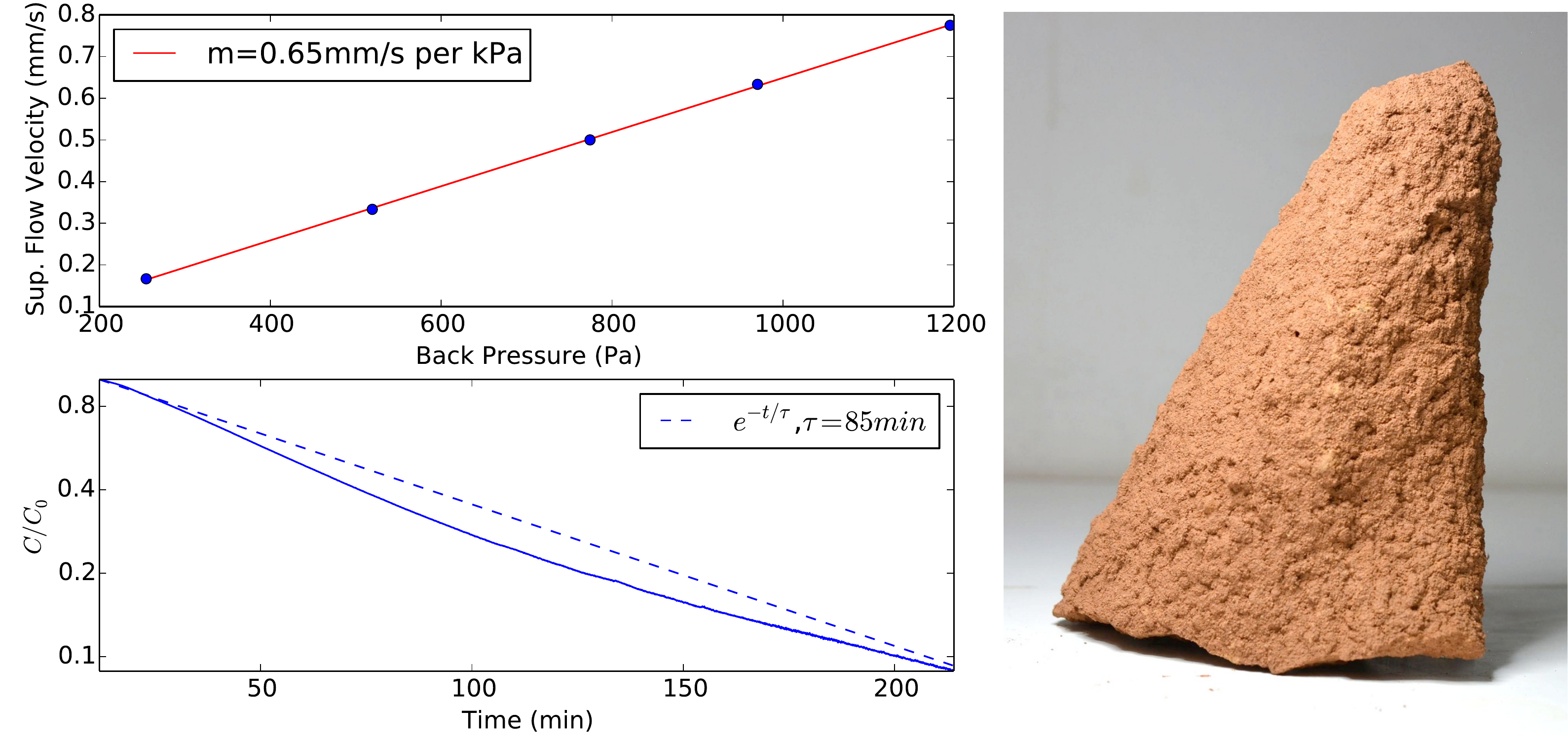}
\caption{Left: A hollow conical sample from a mound flute.  Right: Flow velocity as a function of back pressure measured by sealing the bottom and pulling air into the sample (top), and loss of combustible gas by diffusion in same sample (bottom).}
\label{permdiff}
\end{figure}

A hollow, conical sample of a flute was cut from a mound.  The bottom was sealed with gypsum, such that the pores in the wall material, and length of plastic tubing are the only path in and out of the cone.  Air was pulled by a vacuum pump from the tube through the volumetric flow meter.  The pressure differential between inside and outside of the cone was measured by the displacement of water in a column between the cone and flow meter.  Fig.~\ref{permdiff}(top) shows the 20cm tall conical sample and relationship between back pressure and average flow. 
From this graph, one can read the local flow induced across the mound wall due to a pressure differential from incident wind.  Wind in the area during the study was typically in the range 0-5m/s, which could produce a maximum dynamic pressure $P=\frac{1}{2}\rho v^2 =$ 0-15Pa, giving a maximum flow through the surface of 0.01mm/s.  With even the most liberal approximations, this is not enough to produce bulk flow of the order we've measured, in agreement with the observed negligibly small transient flows in tests with a powerful fan.  If macroscopic holes penetrated the mound surface in some locations, they would dramatically change the permeability estimate of the mound as a whole.  However, such holes were not observed in these mounds, and the species seems to fill in even the smallest holes.  This behavior contrasts that of other species, which appear to tolerate some holes; {\it Odontotermes obesus} actively closed narrow holes made for the $\text{CO}_2$ measurement, {while we have observed that} {\it Macrotermes michaelseni} in Namibia did not.

Impermeability to bulk flow of the wall does not mean non-porous or impermeable to diffusion. Cooking gas was injected into the conical sample and measured by combustible gas sensor that was sealed inside the conical sample.  Fig.~\ref{permdiff}(bottom) shows that it diffuses out the surface over the course of about two hours (following close to exponential decay).
\subsection{$\text{CO}_2$ measurements}
One large ($\sim2m$ tall), apparently healthy mound was chosen (that shown in Fig. 1 (a-b), Fig. 3, and the accompanying video) for measurements.  One hole was drilled from ground level diagonally down into the nest and another into the central chimney 1.5m above ground. 1/4 inch tubing was inserted in the holes and left over night such that the termites sealed the holes at the surface leaving the tubes snugly in place.  A Cozir wide range IR LED $\text{CO}_2$ sensor was fitted with a custom machined, air-tight cap with two nozzles, such that air pulled into the cap would gradually diffuse across the sensor membrane and the response could be recorded with an Arduino onto a laptop computer.  For most of one 24 hour cycle, every 15 minutes air was drawn from each of the tubes in the mound through the sensor with a 50mL syringe, pulling gradually until the response leveled out, meaning the full concentration of mound air had diffused across the sensor membrane.  When termites periodically sealed the end of the tube inside the mound, a few mL of water was forced into the tube, softening and breaking the seal so measurements could continue.


\section{Prediction of Mean Flow Speed}
As a model for a convective circuit within a termite mound, we choose a pipe radius $r$, in the shape of a closed vertical loop of height $h$, where the temperature difference between left and right side of the loop is  $\Delta T$. The total driving pressure is $\rho \alpha \Delta T  g  h$, where $\rho$ is air density and $\alpha$ is the coefficient of thermal expansion,  and Poiseuille's law gives 
\begin{equation}
Q = \underbrace{\rho \alpha \Delta T g  h}_{\text{Driving Pressure}} \cdot \underbrace{\tfrac{\pi r^{4}}{8 \mu h \cdot 2},}_{\text{Poiseuille Resistance}}
\end{equation}
where the factor of two comes from the resistance on both sides of the loop.  Calculating flow speed:
\begin{equation}
V =  \frac{Q}{\pi r^{2}}  = \frac{\rho \alpha \Delta T g  r^{2}}{16 \mu},
\end{equation}
and plugging in values of $\Delta T = 3^{\circ} \text{C}$, $r$ = 3cm, $\mu/\rho$  = .16 cm$^{2}$/s and $\alpha = \frac{1}{300 ^{\circ} \text{C}}$, we obtain a result of $\sim$35 cm/s.  This speed is $\sim$10 times higher than those observed, likely due to oversimplifying internal geometry; disorder and variation in conduit size favors high resistance bottlenecks which reduce the mean flow speed.  The calculation demonstrates that observed thermal gradients and crude dimensions are \textit{sufficient} to produce flow of the order measured.

%
\section{Flow Sensor}
\begin{figure}[h!]
\centering
\includegraphics[height=0.3\columnwidth]{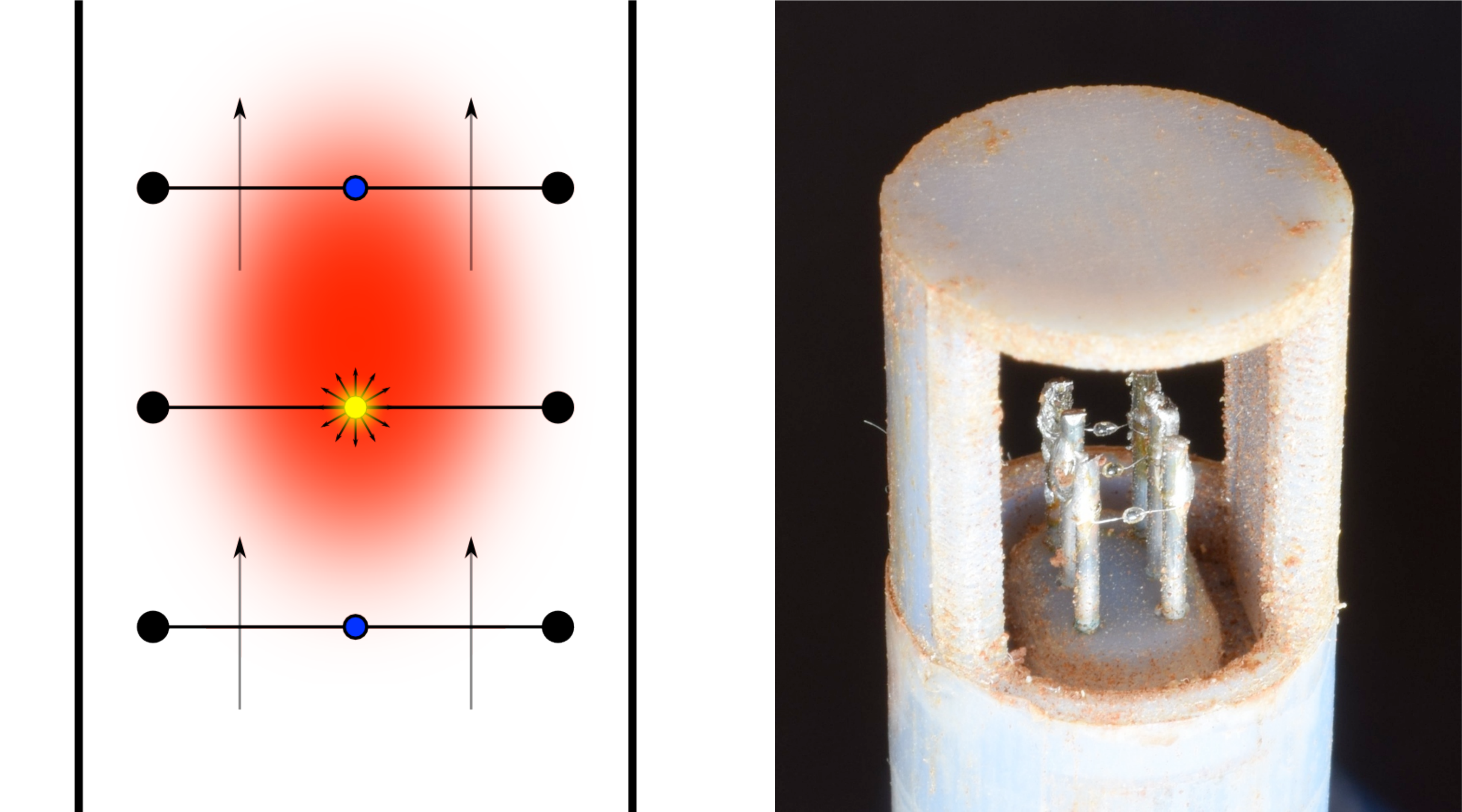}
\caption{\label{beads} Left: Sensor bead schematic.  A brief pulse of heat diffuses towards the nearby beads measuring temperature, with a bias in the direction of air flow.  Right: Image of the sensor head showing three beads aligned with window of cap. }
\end{figure}

Our probe consists of three 0.3mm diameter glass coated thermistor beads (Victory engineering corporation, NJ, $R_0=20k\Omega$) held exposed by fine leads in a line with 2.5mm spacing.  The center bead is used as a heat source, either pulsed or continuous, in steady and transient modes, respectively.  As the heat diffuses outward, its bias depends on the direction and magnitude of flow through the sensor, as depicted in Fig. \ref{beads}(Left).  The operating principle is similar to that of some sensitive pulsed wire anemometers~\cite{pulse-wire}.  Flow is quantified by comparing the signals (effectively temperatures) from either neighboring bead.

The plastic housing for the sensor was drawn with Solidworks and printed on a Object Connex500 3D printer and the individual thermistors were connected to the electronics via shielded Cat 7 ethernet cable.  The thermistor beads can be seen in the large window in the protective cap of the sensor in Fig. \ref{beads}(Right).

\begin{figure}[h!]
\centering
\includegraphics[height=0.45\columnwidth]{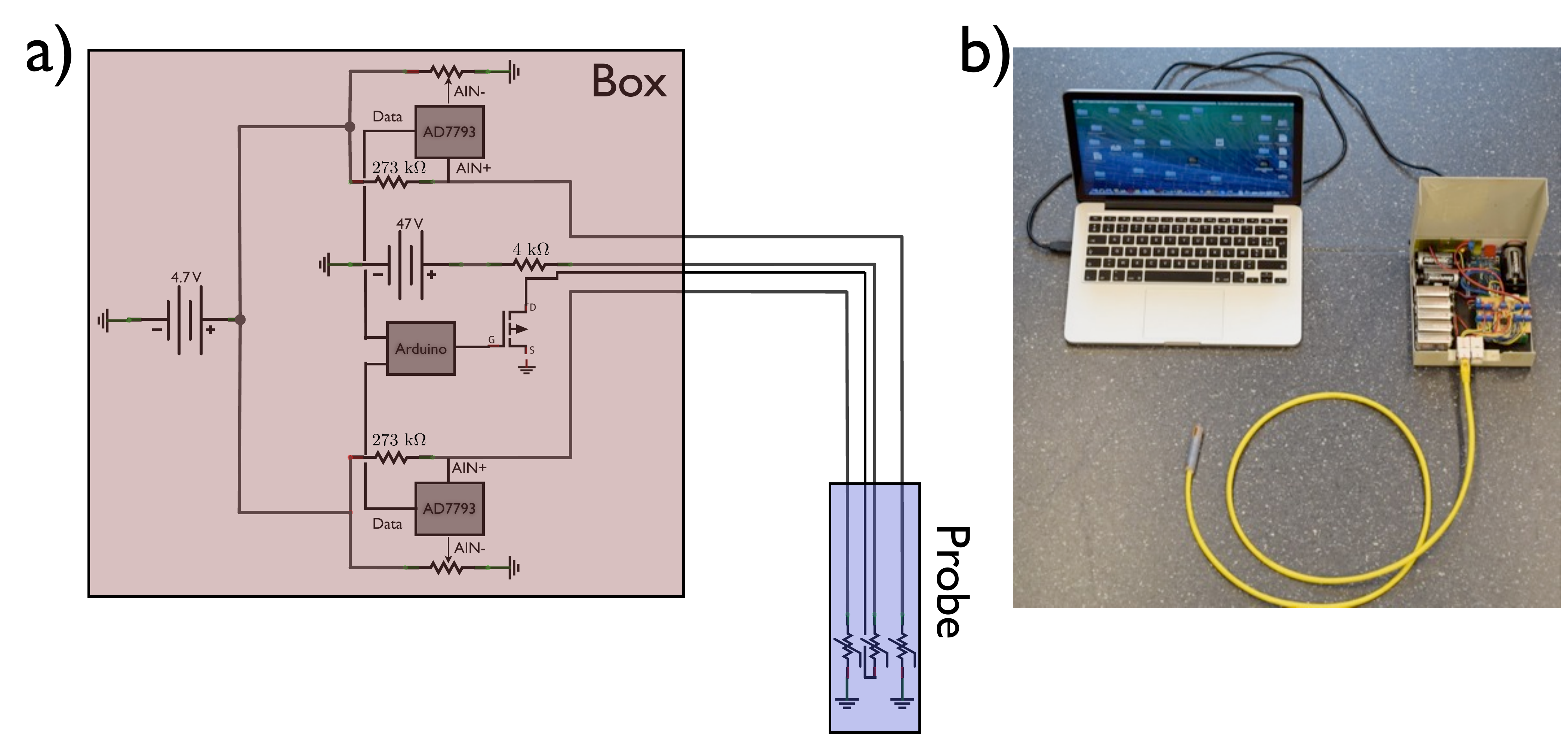}
\caption{\label{Circuit} a) Schematic of circuit. b) Photograph of setup. The sensor is connected by ethernet cable to the box containing supplemental electronics and Arduino, which is connected to a computer by USB.}
\end{figure}

\subsection{Electronics}
Data was collected and the sensor was controlled via an Arduino Uno, while connected to a laptop computer using custom scripts written in Processing.  To improve the resolution of the tiny temperature signals, the voltages are measured with AD7793 (Analog Devices) ADCs recording at 10Hz, and a stable baseline signal was supplied by three C batteries in series.  The pulsing voltage was supplied by 5 9V batteries, connected momentarily to the middle bead by opening a SIHLZ14 (Vishay) MOSFET.  Trimmer potentiometers were used to keep the baseline temperature signal within range of the ADCs.  The basic circuit diagram can be seen in Fig.~\ref{Circuit}(a).  The supplemental electronics, batteries, and Arduino were housed in a portable metal box, as shown in Fig.~\ref{Circuit}(b).

\begin{figure}[h!]
\centering
\includegraphics[width=\columnwidth]{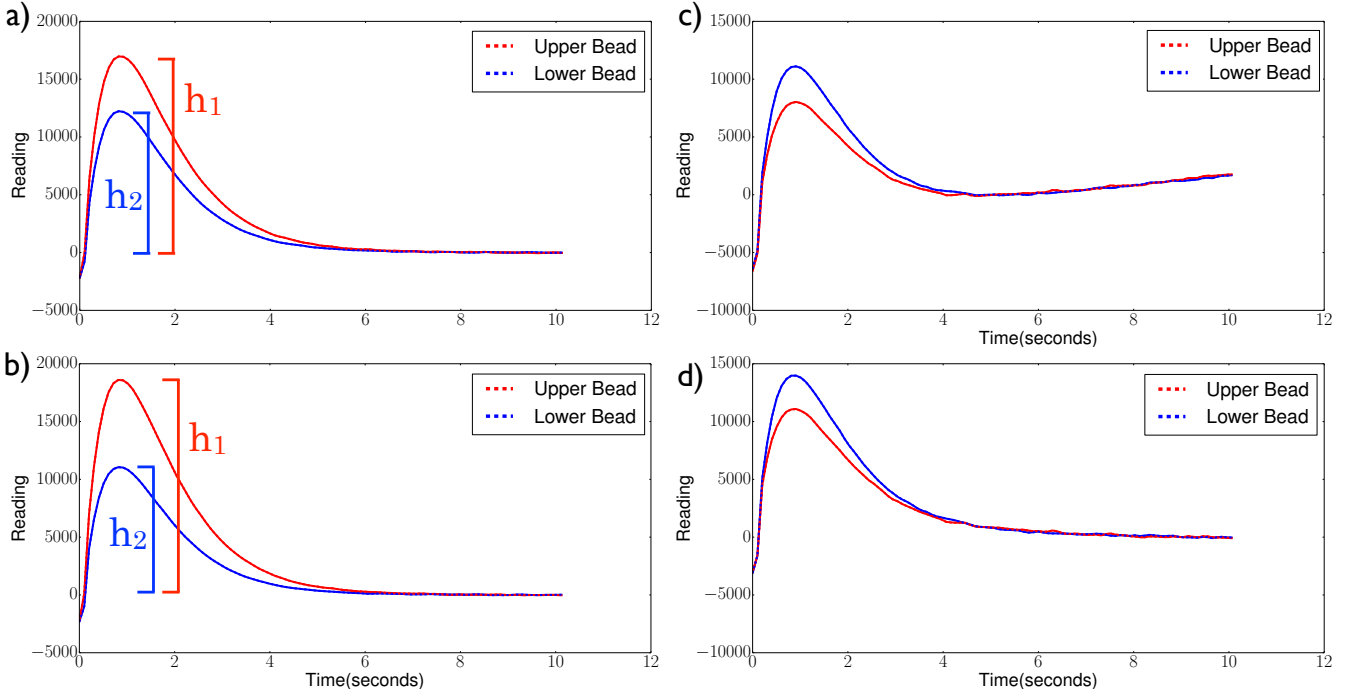}
\caption{\label{response}Averaged responses from two thermistors as heat bolus diffuses away from center bead for a) Zero flow b) 1.4 cm/s upwards flow in calibration tube; c) Before and d) after subtraction of temperature drift in field data.}
\end{figure}

\subsection{Steady mode function}

In order to obtain a passive measurement of steady (average) flow, the center bead, in series with 4k$\Omega$ was pulsed for 30 ms with 47V every 10 seconds.  We estimate its temperature to briefly reach $180-300^{\circ}$C.  This produced a small bolus of warm air that diffused outward toward neighboring beads in either direction.  The tiny change in resistance due to the expanding heat bolus was measured in the neighboring beads, as shown in Fig. \ref{response}.  In the absence of flow through the sensor, there is a slight upward bias in the response magnitudes due to thermal buoyancy of the bolus.  Upward and downward flow biases the relative responses of the two thermistors in a reproducible way.  The log of the ratio of the maxima ln$(h_{1}/h_{2})$ was used as a metric to calculate flow velocity (see steady flow calibration section).  For experimental data sets, temperature drift must be subtracted as the probe equilibrates with the interior flute temperature.

\begin{figure}[h!]
\centering
\includegraphics[width=0.75\columnwidth]{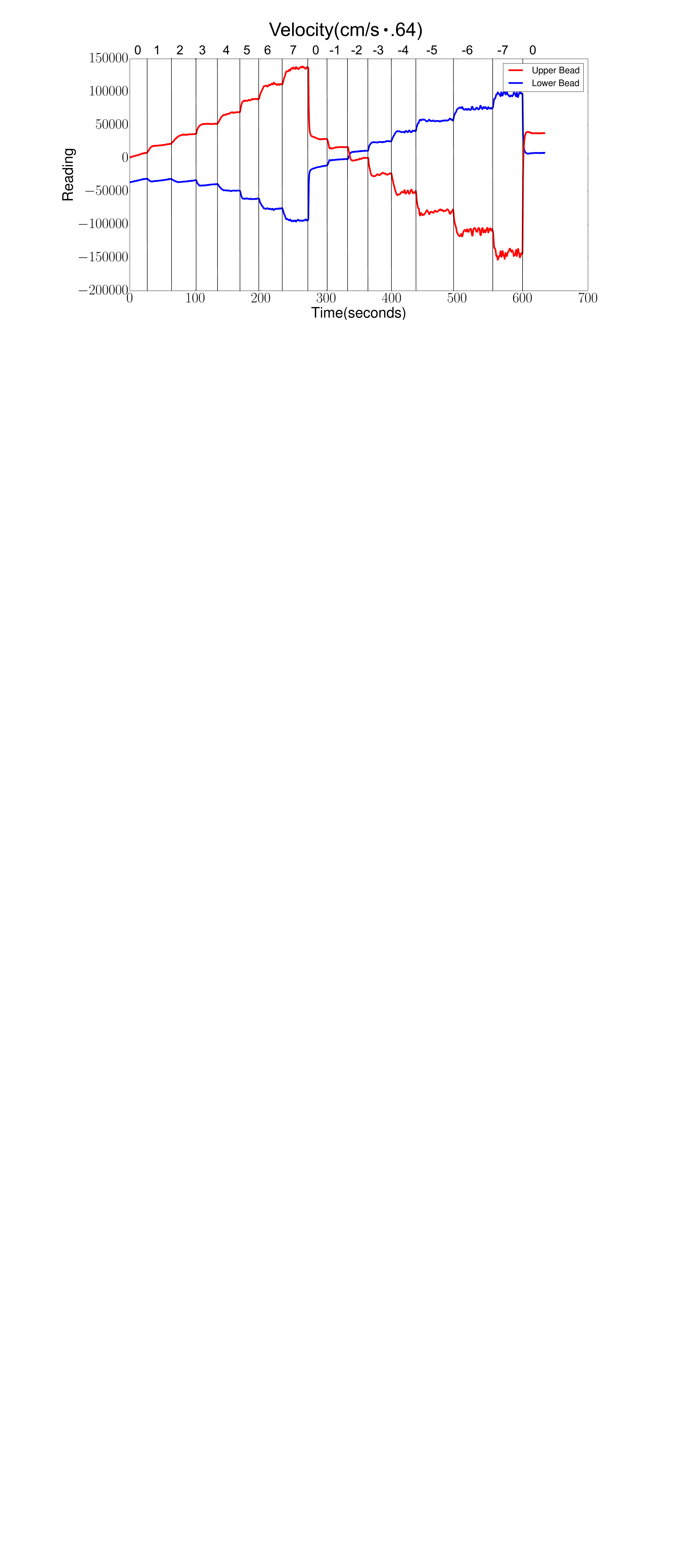}
\caption{\label{transient}Relative readings upper and lower beads in transient mode as flow was varied in increments of .64cm/s}
\end{figure}

\subsection{Transient mode function}

While in steady mode, infrequent, brief pulsing of heat avoids troublesome induced convective currents, it prohibits measurement of high frequency($\gtrsim .1$ Hz) changes in flow.  In transient mode of the same sensor, continuous voltage 47 V is applied across the middle bead, such that transients in the flows can be measured from the instantaneous ratio of responses in the neighboring beads.  The trade-off is that information about the baseline flow is lost, because induced currents, which depend on several parameters, including unknown details of tunnel geometry, can dominate the signal.  Fig. \ref{transient} shows the response of the neighboring beads in transient mode for {some prescribed transients. At $\sim0$ cm/s the dependence on the difference between the two readings is the weakest($\sim\frac{30,000}{cm/s}$); assuming that slope, the largest fluctuation observed in the field corresponded to $\sim1 mm/s$}.  

\subsection{Steady flow calibration}
\begin{figure}[h!]
\centering
\includegraphics[width=0.5\columnwidth]{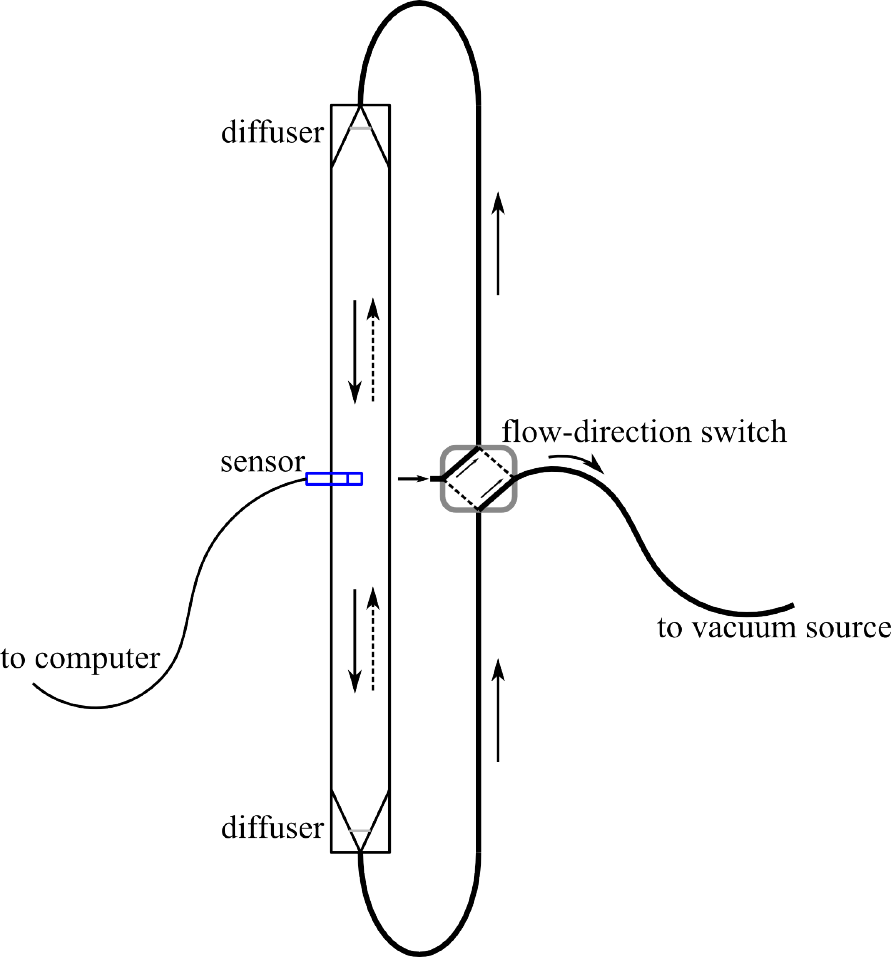}
\caption{\label{tubescheme}Schematic of tube used for calibration.  Arrows show direction path of air from inlet toward vacuum source.}
\end{figure}

A 5.6 cm diameter, 2m long, vertical acrylic tube was used as a sample conduit for calibration of the sensor (see Fig.~\ref{tubescheme}).  The tube was covered with a layer of thermal insulation to prevent external thermal gradients from influencing flow, which was observed when the bottom of the tube was exposed to the sun through a lab window.  Air was pulled through the tube by a vacuum source and mass flow controller (Alicat MC-20slpm) in series.  To prevent heterogeneous inertial flows where the narrow connector tube met the wider conduit, air was directed through a fine metal mesh before entering a conical flow rectifier.  The whole set-up was up-down symmetric, such that by rotating two valves, flow direction could be reversed without changing the setup geometry.  The sensor was inserted into a hole in the middle of the tube and oriented so that the beads were aligned with the tube. 

\begin{figure}[h!]
\centering
\includegraphics[width=.65\columnwidth]{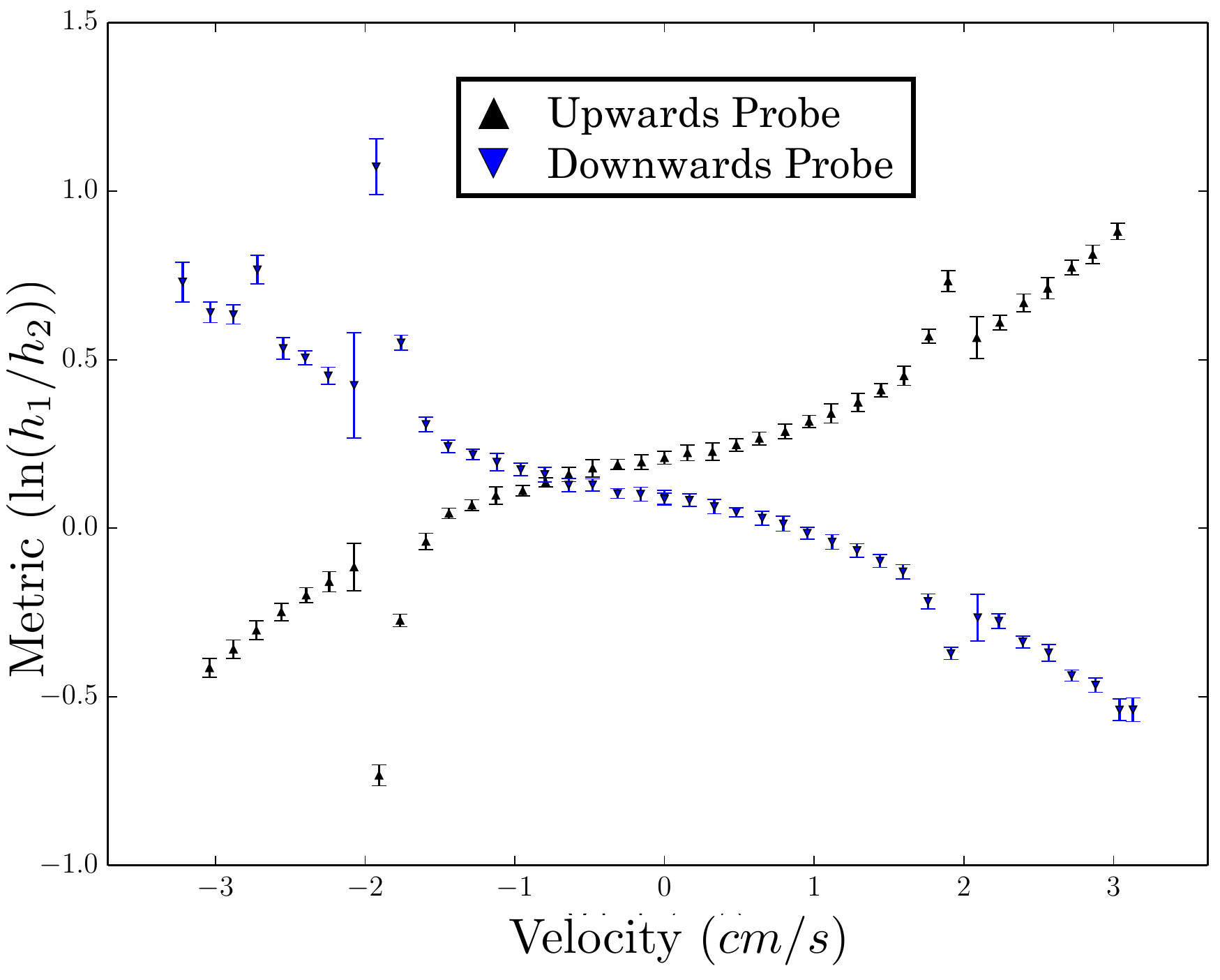}
\caption{\label{calibration} Calibration curves for steady flow sensor measured in the plastic tube. The upwards {black triangles} represent the upwards orientation of the sensor, while the downwards blue triangles represent the downwards orientation. {Error bars are the pulse-to-pulse deviation. }} 
\end{figure}

\begin{figure}[h!]
\centering
\includegraphics[width=.75\columnwidth]{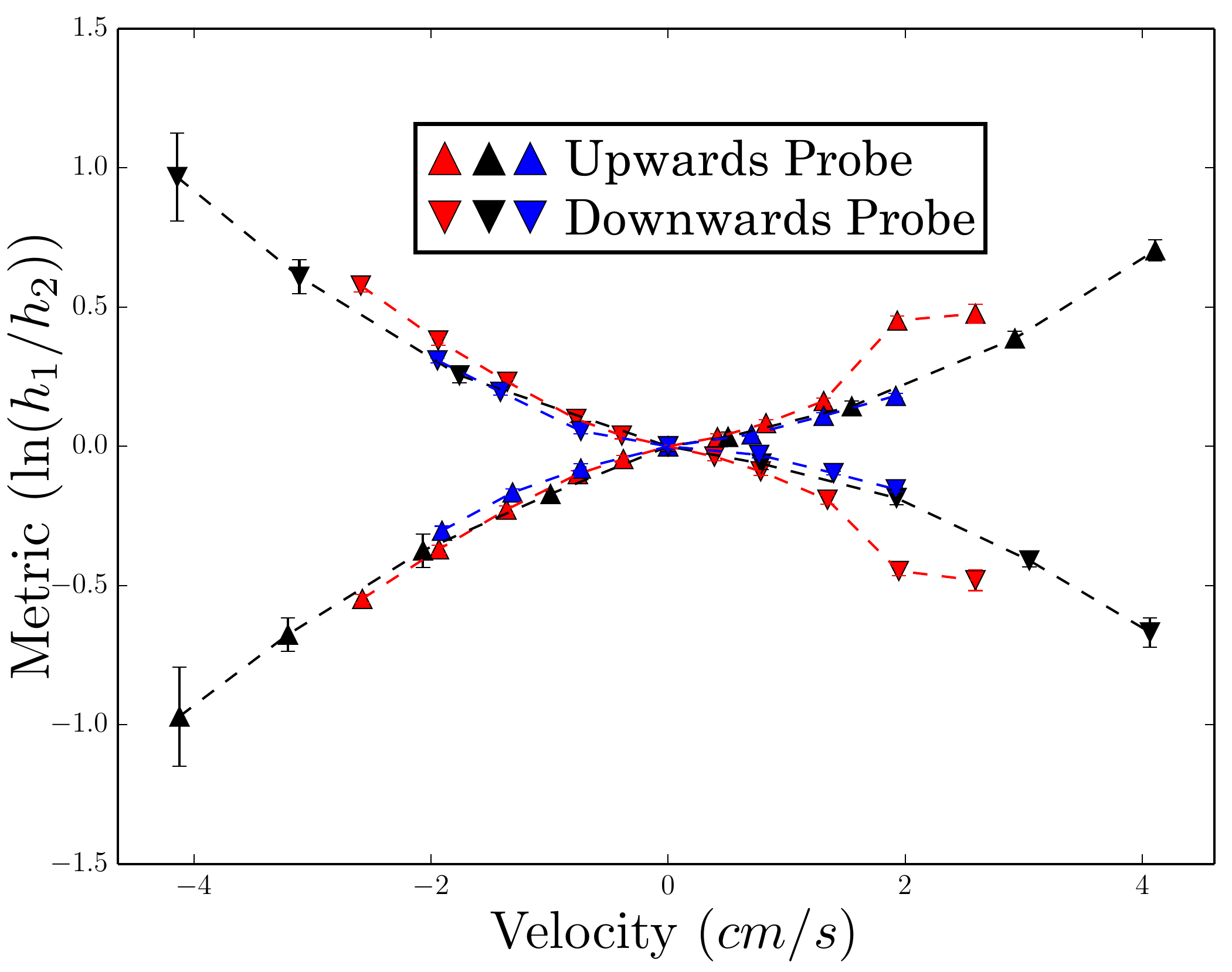}
\caption{\label{repeatability}Shifted calibration curves for three instances of the sensor reveal a predictable relationship between metric and velocity.}
\end{figure}

{As a consistency check and to eliminate possible systematic biases in the probe, every airflow measurement is taken twice using the same probe in two orientations. In one the (arbitrarily named) first bead is above the middle thermistor('upward'); in the other it is directly below ('downward'). To do so, we need separate calibration tables for the 'upward' and 'downward' orientations. }


Fig. \ref{calibration} shows the dependence of the metric on flow velocity in the calibration tube.  The typical 
standard deviation between pulses during calibration is small and does not represent the dominant source of error in the field, where variation in conduit geometry plays a larger role (see Experimental Procedure).  {There is an unusual, though robust, feature at $\sim \pm 2cm/s$, where the response briefly jumps.  This is most likely an effect due to shifting from a regime of viscously dominated laminar flow to inertial dominated laminar flow. We note that this leads to a small range of flow velocities where we overestimate the flow speed(but not the sign).}

{Fig.~\ref{repeatability} shows several calibration curves for the different instances of the sensor, either a spare duplicate sensor, or the same physical sensor after thermistor beads were replaced.  Each instance requires a new calibration, as the the magnitudes of bead responses varies according to the manufacturer's 25\% tolerance, but upon shifting the vertical offset and slope, on can see the qualitative behavior of each instance is the same.}

\subsection{Temperature and humidity dependence}

\begin{figure}[h!]
\centering
\includegraphics[width=0.75\columnwidth]{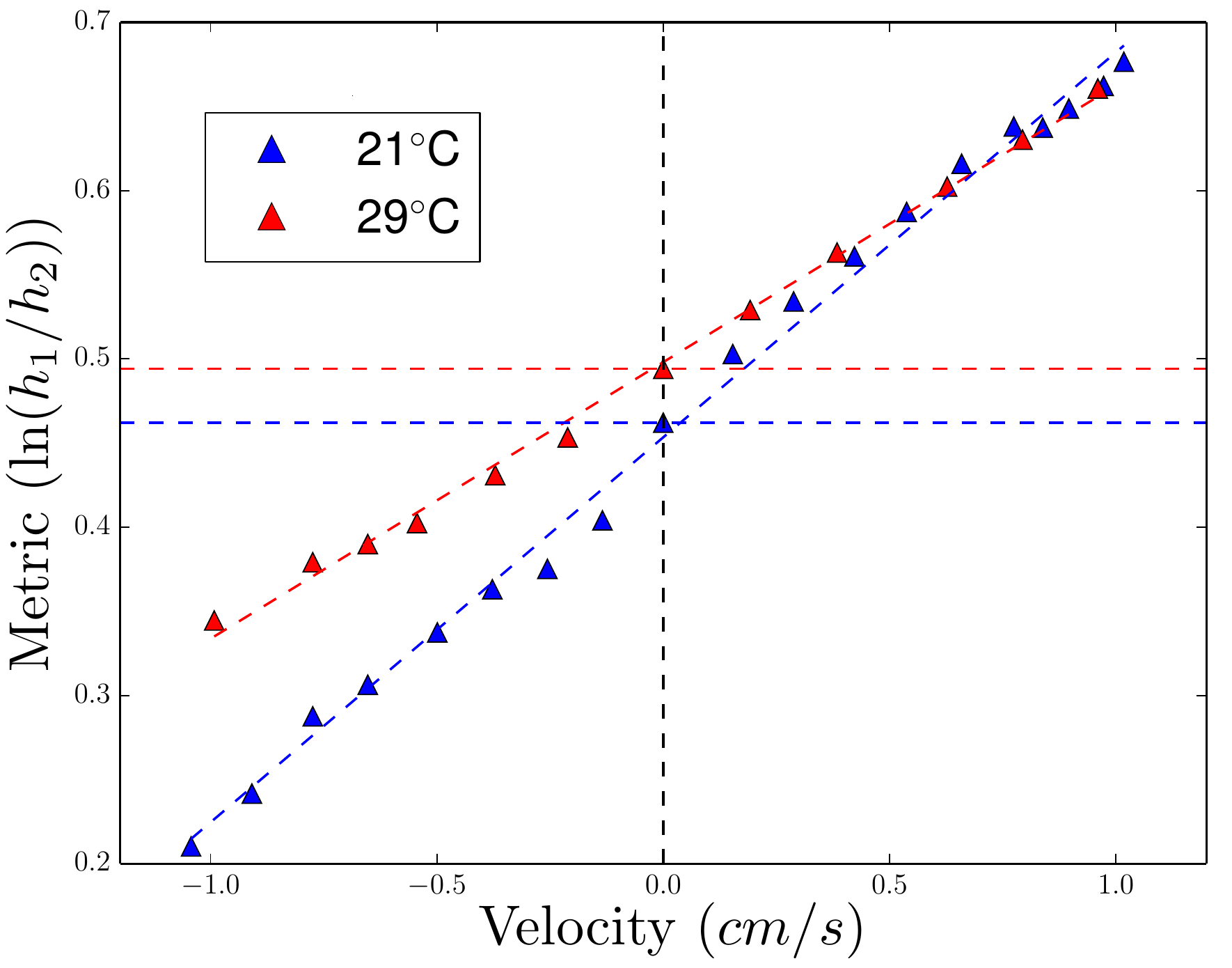}
\caption{\label{tempcurv} {Calibration curves for two temperatures show a slight shift($\sim 0.2 cm/s$) and slope change ({$\sim 40 \%$) for this temperature difference, which is closely matched with the total range observed in live mounds. Horizontal and vertical lines are to guide the eye. }}
}
\end{figure}
\begin{figure}[h!]
\centering
\includegraphics[width=.6\columnwidth]{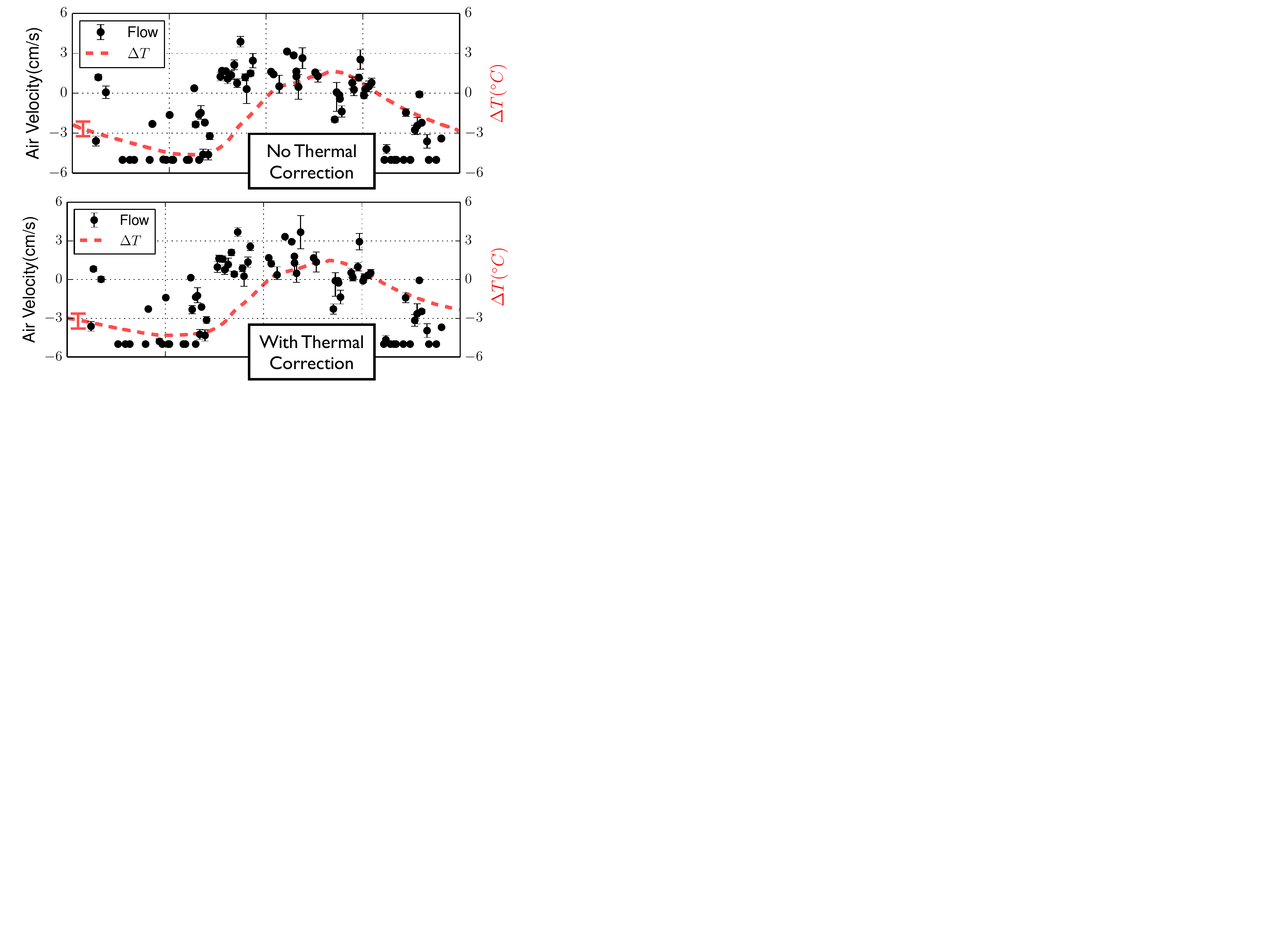}
\caption{\label{thermcorrect} Comparison of calculated velocities where temperature-dependent sensitivity is taken into account.}
\end{figure}
While the calibration curve in laboratory conditions is robust, additional tests were performed to make sure that our metric was not terribly sensitive to other factors which can vary in the field.  Two such factors which should influence the thermal response of the sensor, and therefore its performance are background temperature and humidity.  Fig.~\ref{tempcurv} shows the calibration curve for two different ambient temperatures, which roughly represent the range of temperature in the mound.  Though the curve has a lower slope at higher temperature, which indicates that the sensor underestimates hotter flows, this effect is small compared to the variation of the daily mound flow schedule measured in the field, and would not account for the sign change.  The metric, continuously measured for constant flow was unaffected by a change in ambient relative humidity from 25$\%$ to 70$\%$ (at 29$^{\circ} $C).

{We can estimate this effect by applying a temperature-dependent calibration, where sensitivity is assumed to vary linearly with temperature, interpolated between the slope and offset of the $21^{\circ}$ and  $29^{\circ}$ curves. Using this with the temperature information from inside the mound, the effect of temperature on all live mound values can be approximated, as shown in Fig.~\ref{thermcorrect}.  We can see the values are only slightly modified, causing a tiny enhancement to the trend, where daytime flows become slightly more positive.}
{\subsection{Orientational Dependence}}
\begin{figure}[h!]
\centering
\includegraphics[width=0.75\columnwidth]{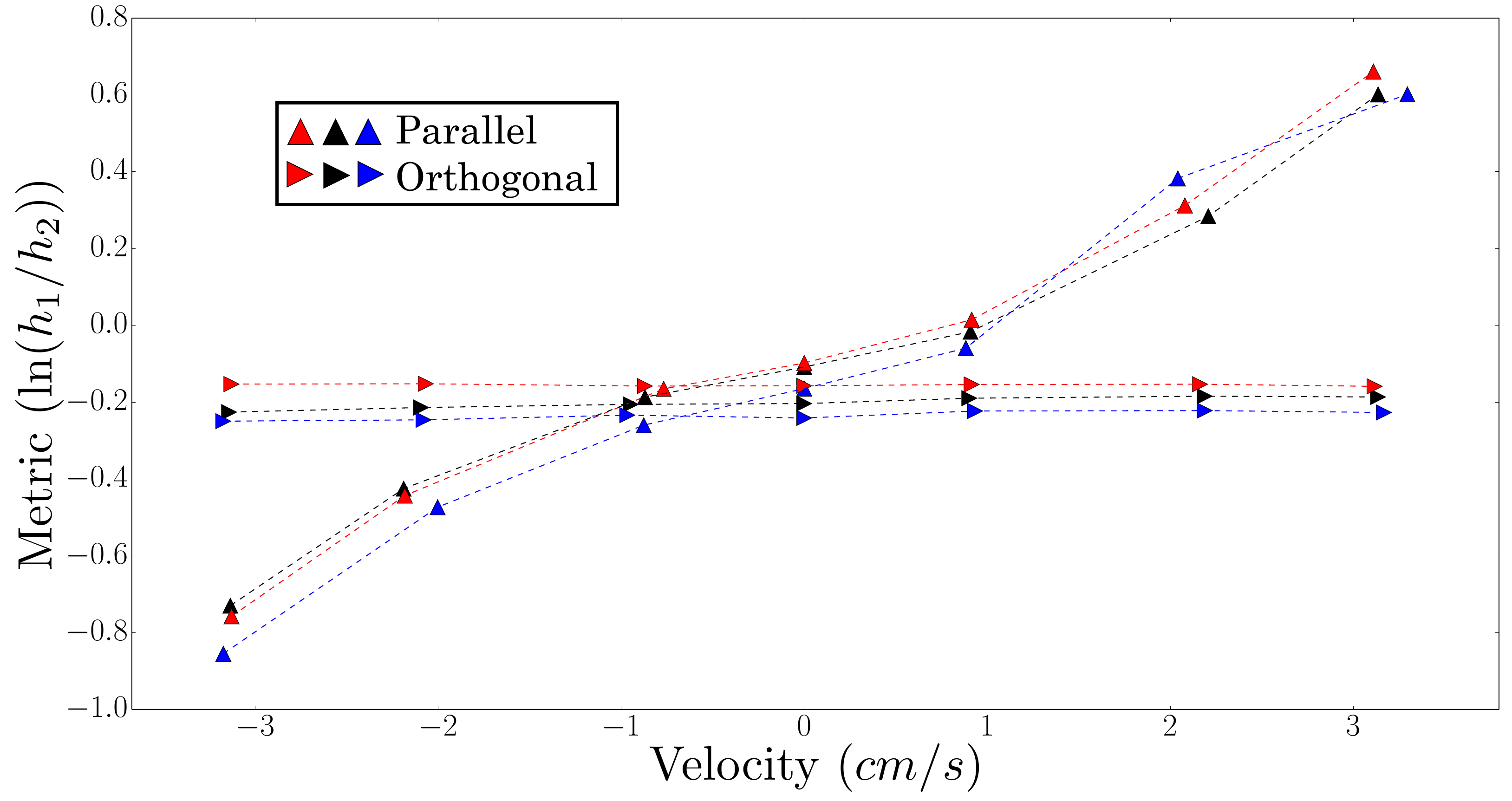}
\caption{\label{aligned_misaligned}{Velocity metric as a function of flow speed tube orientation and relative probe orientation. Regardless of whether the tube is vertical (red), diagonal ($45^{\circ}$, black) or horizontal (blue), the probe is only sensitive to flow velocity when the probe is aligned parallel with the tube (upwards triangles), and not when orthogonal (rightwards triangles)}}
\label{SD1}
\end{figure}
\begin{figure}[h!]
\centering
\includegraphics[width=.75\columnwidth]{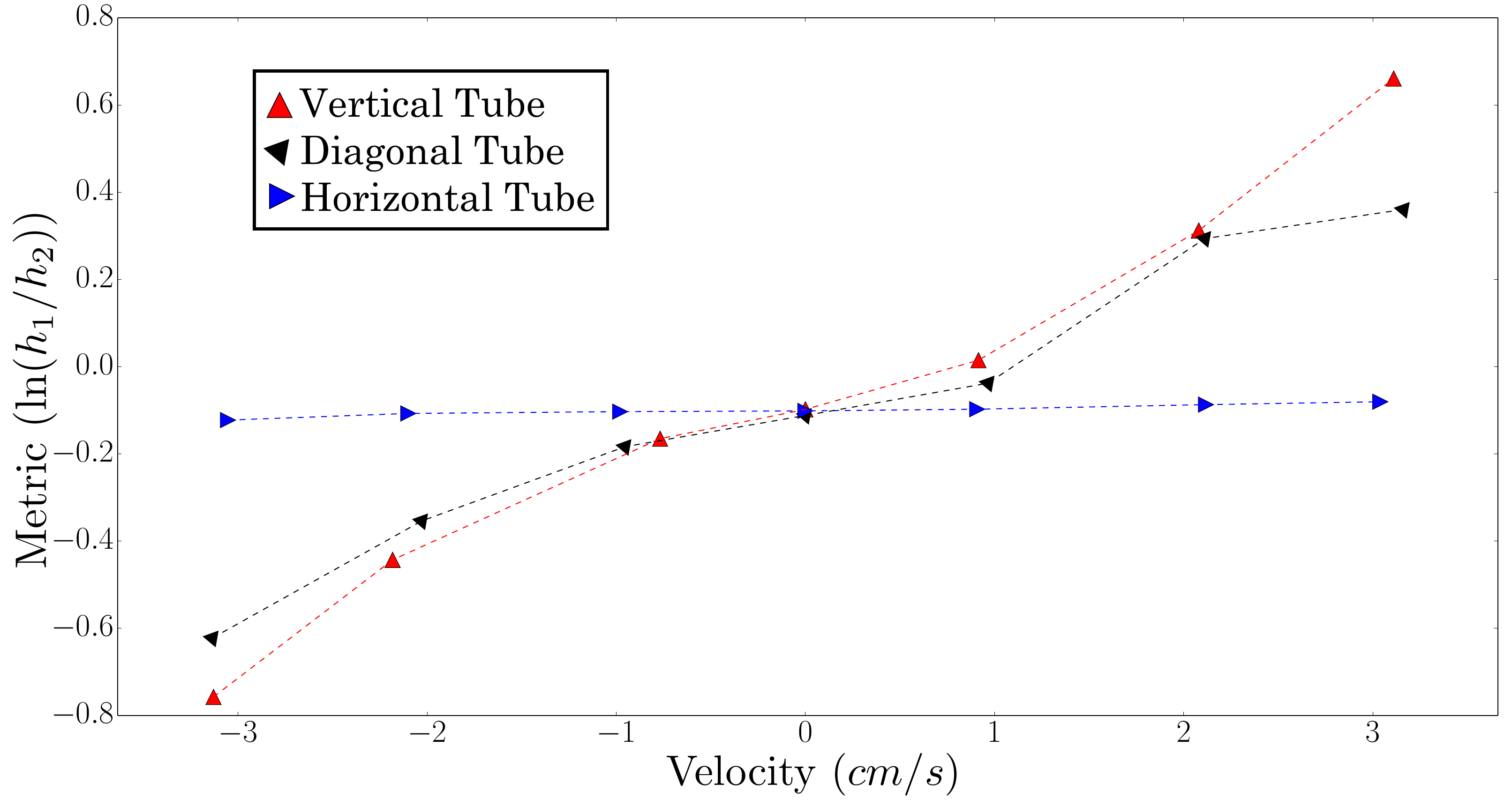}
\caption{\label{tube_tilt_test}{Velocity metric as a function of flow speed and tube orientation for a vertical probe. When the tube is vertical($90^{\circ}$), sensitivity is highest, diagonal($45^{\circ}$), gives reduced sensitivity, and the probe is not sensitive to horizontal flows($90^{\circ}$) at all.}}
\label{SD1}
\end{figure}

{
As mentioned above, the sensor is calibrated exclusively for vertical flow.  Though it was not difficult to ensure that local conduit orientation was vertical where the sensor was placed, and that flow was therefore predominantly vertical, it is not unreasonable to suspect that a horizontal flow component could be misinterpreted by the sensor.}
\begin{figure}[h!]
\centering
\includegraphics[width=.75\columnwidth]{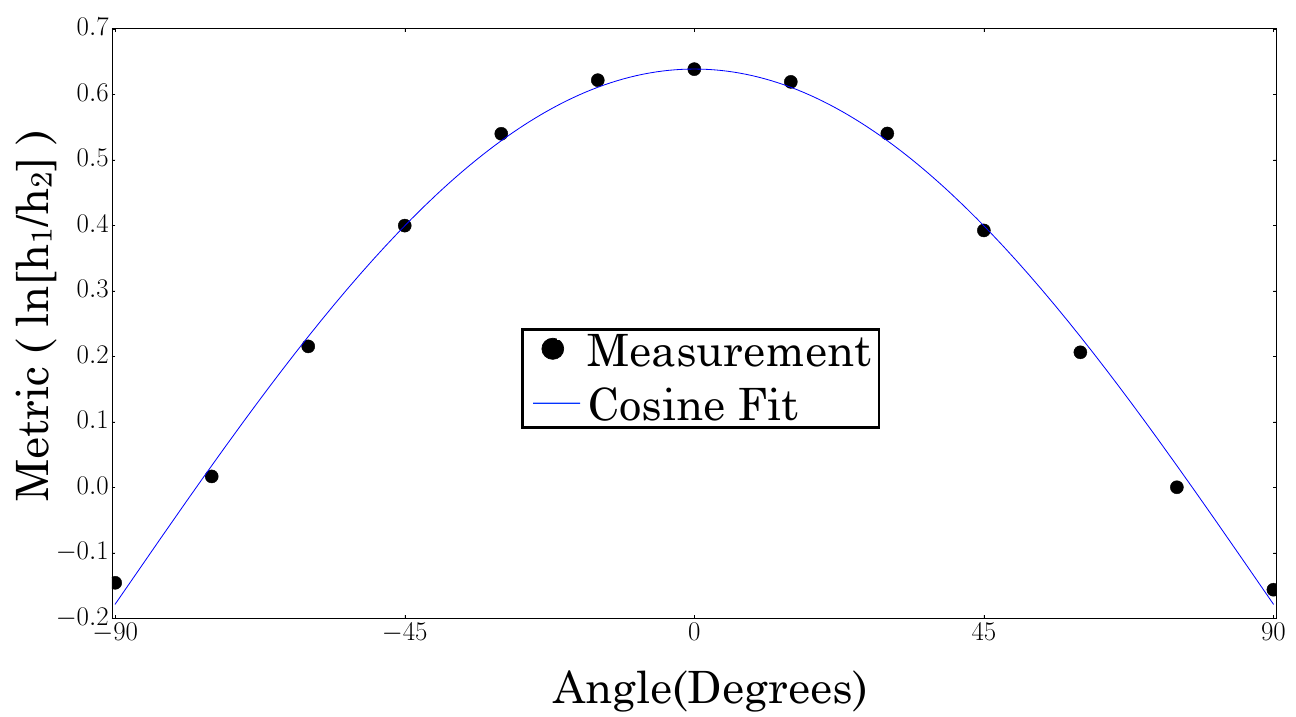}
\caption{\label{slow_tilt_test}{Velocity metric as a function of probe rotation in a tube with upwards flow.}}
\label{SD1}
\end{figure}

{Fig.~\ref{aligned_misaligned} shows the metric as a function of flow velocity for 3 orientations of the calibration tube; horizontal, 45$^{\circ}$, and vertical, when the sensor is either parallel or orthogonal (cap windows 90$^{\circ}$ from tube axis).  Negligible change in metric for all orientations in which the sensor was misaligned indicates that the sensor cap effectively rectifies flow for all orientations; any flow component perpendicular to the sensor would have negligible effect on the measured value.  When the flow is aligned with the sensor, for the full range of probe and tube orientations, there is only a small shift in the curve.  Additionally, Fig. \ref{tube_tilt_test} shows that a vertically placed sensor's sensitivity to flow decreases to zero as the direction of flow is changed from vertical to horizontal. These rule out the possibility of non-trivial sensitivity to conduit orientation.}

{Fig.~\ref{slow_tilt_test} shows the metric as a function of probe angle in a vertical tube of constant flow velocity 3.1 cm/s, where the probe is aligned ($0^{\circ}$) or orthogonal ($\pm 90^{\circ}$) to the tube. The good fit to a cosine shows that flows not aligned with the probe (horizontal flows in the field) do not significantly affect the measurement. }

\section{Additional Measurements: Long-term flow in a dead mound}

\begin{figure}[h!]
\centering
\includegraphics[width=.75\columnwidth]{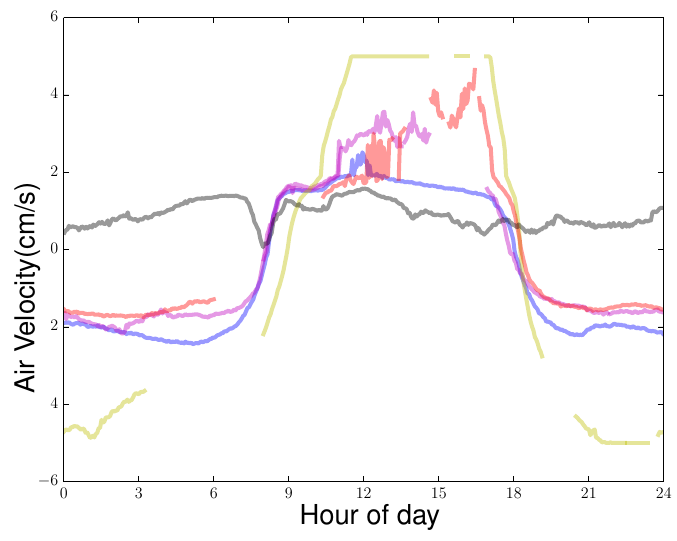}
\caption{Continuous flow measurements of 5 flutes in a dead mound.}
\label{SD1}
\end{figure}

Continuous flow in five flutes from the dead mound were measured, shown in Fig.~\ref{SD1}.  Four out of five flutes follow the observed trend.  The heterogeneity is likely due to the complex geometry and is consistent with the few live mound measurements which also go against the trend.

\begin{figure}[h!]
\centering
\includegraphics[width=0.75\columnwidth]{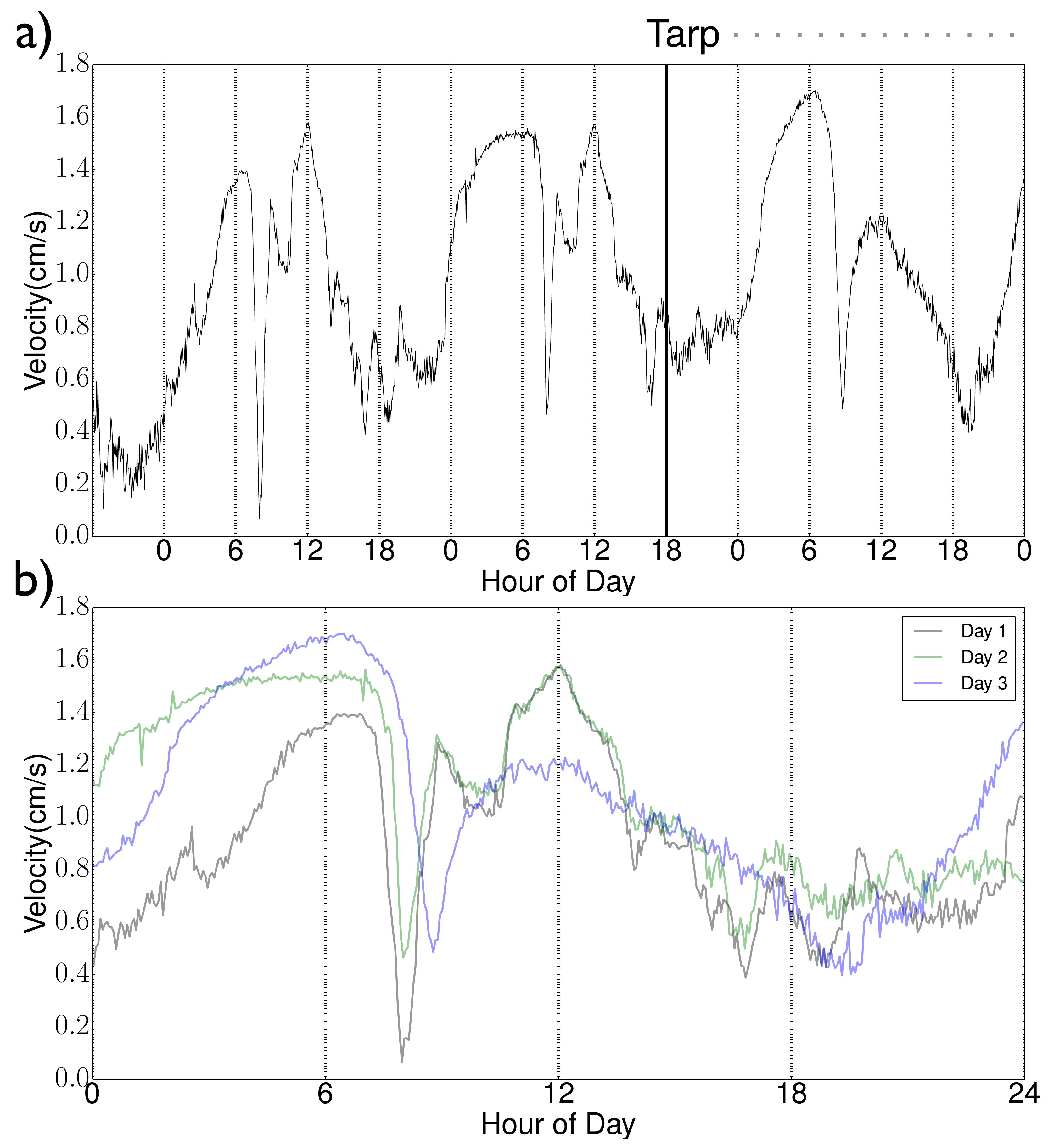}
\caption{a) 3 day flute flow measurement. A tarp was hung to shade the mound on the third day. b) The same data, with days 1, 2, 3 overlaid on top of each other.}
\label{SD2}
\end{figure} 

One flute (which happened to go against the observed trend) was measured for three days. For the first two days, the mound (at the edge of a forest) was exposed to partial direct sunlight.  It can be seen in Fig.~\ref{SD2} that the measured air velocity strictly follows a daily schedule, where fine features repeat themselves. On the third day, a tarp was positioned above the mound, keeping the mound shaded from any direct sunlight, but exposed to ambient temperatures. Some midday features are missing, presumably those directly induced by solar heating, but the general trend remains the same.

{
\section{Permeability Measurements}
\begin{figure}[h!]
\centering
\includegraphics[width= .7\columnwidth]{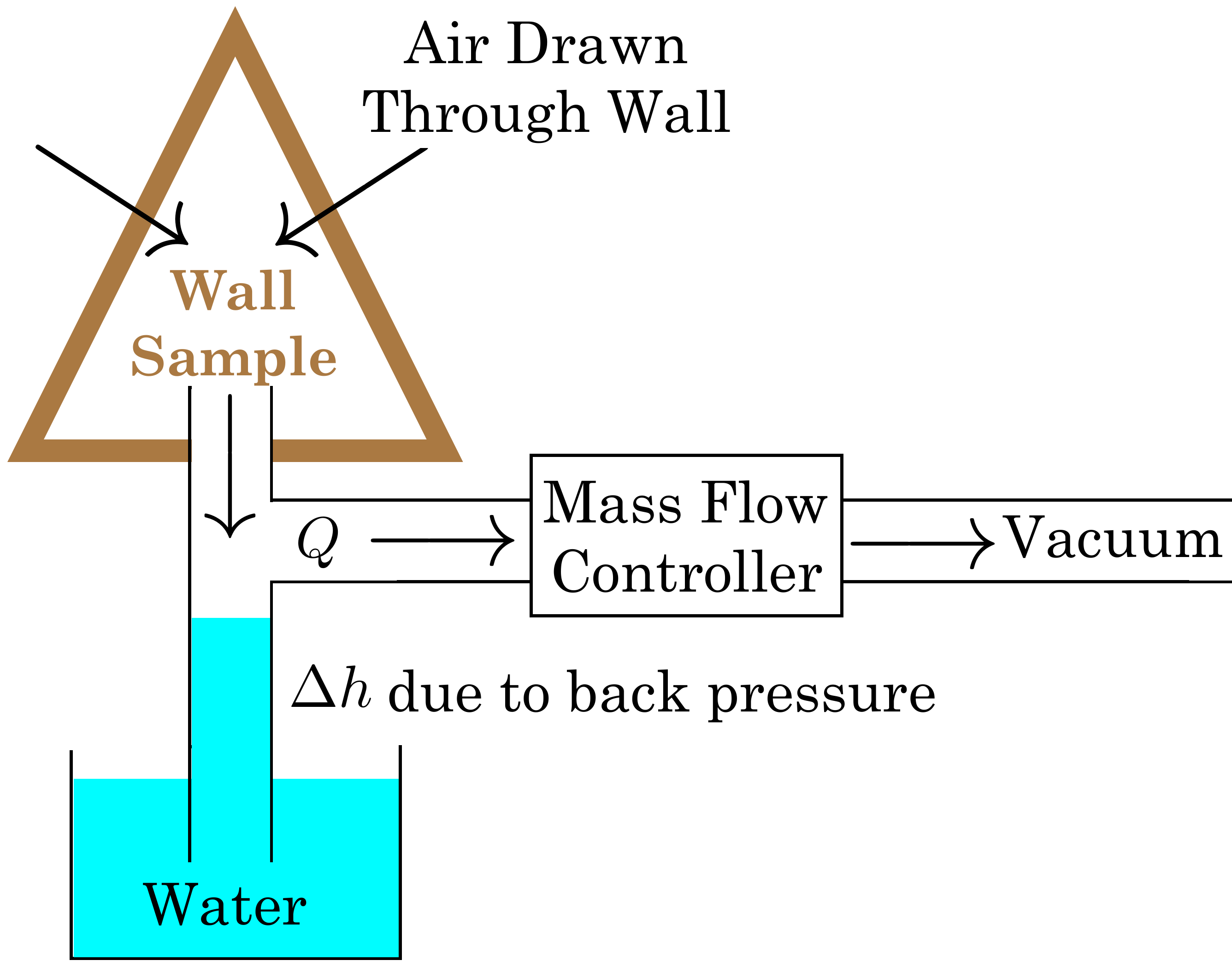}
\caption{Schematic of permeability measurement apparatus.}
\label{SD2}
\end{figure} 
The mound material is {37$-$47\% air by volume}, and has an average pore size of $\sim 5 \mu m$, roughly the mean particle size \cite{tejas}. To determine the degree to which this admits bulk flow of air, air was pulled through the tube by a vacuum source, mass flow controller (Alicat MC-20slpm), and conical sample of mound wall sealed with gypsum. In parallel was a piece of glassware inside a water reservoir, connected to the mound sample on top (Fig.~\ref{SD2}). {The height differential $\Delta h$ allows us to calculate the {back pressure $\Delta P$ = $g \rho \Delta h$.} This, combined with the measured air flow $Q$ (measured by the mass flow controller) gives the permeability of mound sample $\kappa = Q/(\Delta P  A)$, where $A$ is the area of the sample. Our measured values for a larger mound sample are in rough agreement with the measurements \cite{tejas} on hydraulic conductivity.}
}

\end{document}